\documentstyle[12pt,psfig]{l-aa}

\begin{document}
\thesaurus{03(11.04.1; 11.03.1; 11.03.4 A548; 11.03.4 A3367) } 
\title{A redshift survey between the clusters of galaxies A548 and A3367
       \thanks{based on observations collected at the European
               Southern Observatory, La Silla, Chile}}
%
\author{G. Andreuzzi \inst{1,2} \and S. Bardelli \inst{3} \and
R. Scaramella \inst{2}  \and E. Zucca \inst{4}}
\offprints{S. Bardelli (bardelli@astrts.oat.ts.astro.it) }
\institute{
Osservatorio Astronomico di Capodimonte, via Moiariello 16, I-80131, 
Napoli,  Italy
\and 
Osservatorio Astronomico di Roma,  via Osservatorio 2, 
 I-00040, Monteporzio Catone (RM), Italy
\and 
Osservatorio Astronomico di Trieste, via G.B. Tiepolo 11, I-34131,
Trieste,  Italy
\and 
Osservatorio Astronomico di Bologna, via Zamboni 33, I-40126, 
Bologna, Italy}
\date{Received ; accepted }
\maketitle

\begin{abstract}

In this paper we present the results of a spectroscopic survey of
galaxies in an area between the two clusters of galaxies A548 and
A3367, suspected to be a close and interacting pair.  With the use of
multifiber spectroscopy, we measured 180 new velocities of galaxies in
the central part of A3367 and in the external regions of A548.

The redshift histogram shows the presence of three velocity peaks, at
$v\sim 12000$ km/s, $v\sim 30000$ km/s and $v\sim 40000$ km/s,
respectively.  For these we estimate the density excess, the mean
velocity, and the velocity dispersion.

The first clump corresponds to an elongation of A548: in particular we
found a correspondence between the features of this peak and the
substructures of A548.  The second peak has a velocity dispersion
which is typical of clusters and the distribution of its members on
the plane of the sky corresponds to the highest density peak in A3367.
We therefore suggest that the name A3367 has to be attributed to this
clump.

Our general conclusion is that, differently from expected, A548 and
A3367 do not form a close pair of merging clusters, since the two
structures are at significantly different redshifts.  Moreover, we
found that the complex dynamical structure of A548 has large
coherence, with a projected extension in the range of 1-3 h$^{-1}$
Mpc.

\keywords{ Galaxies: clusters: individual: A548, A3367; Galaxies: clusters of;
           Galaxies: distances and redshifts }
\end{abstract}
%
%
\section{Introduction}

The determination of the galaxy distribution in clusters provides
information on the status and the history of these structures through
the study of their dynamics.  Detailed studies spanning the entire
range of morphologies of rich clusters of galaxies are important for
understanding the formation and evolution of these systems.  In a
class of current cosmological models (e.g. cold dark matter dominated),
rich clusters are formed hierarchically, by accretion of smaller
subunits.

Several clusters are indeed known to present very lumpy morphologies
(see e.g.  Kriessler \& Beers 1997 and references therein) revealing
that these systems are in a merging process. The best studied examples
are A2256 (Briel, Henry \& B\"ohringer 1992), where a small group is
detected in the X-ray band nearby the cluster center, and Coma, where
a number of substructures are revealed (Biviano et al. 1996).  Among
the most spectacular cases are the encounters between clusters of
similar richness, as for the A3558 complex (Bardelli et al. 1994,
1996, 1998a, 1998b), where the dynamical processes reach unusual
intensities, or the cluster A3528, which is actually split into two
merging X-ray emitting regions of similar properties (Schindler 1996).

The study of merging clusters is important because this process is
thought to be responsible for a wide number of properties of the
cluster galaxy population.  Radio halos and relicts of radiosources are
found in clusters that visually present some degree of disturbance
(Feretti \& Giovannini 1996)
and Burns et al. (1994) explained as a consequence of a merging
event the presence of post-starburst galaxies in the large scale
X--ray emitting filament connecting Coma with the NGC4839 group.

A good starting point to individuate merging cluster candidates is
that to extract close pairs from supercluster catalogues, as f.i. the
list of Zucca et al. (1993), which reports groups of ACO clusters
(Abell, Corwin \& Olowin 1989) as a function of the density contrast.
In this catalogue, the cluster pairs individuated by a density excess
$>200$ are very close systems, where often the nuclei are separated by
less than one Abell radius ($\sim 1.5$ h$^{-1}$ Mpc, hereafter
h=H$_o$/100): one of these pairs is formed by A548 and A3367.

The centers of the two clusters in the ACO catalog are separated on
the plane of the sky by 77 arcmin, corresponding to $\sim 2$ h$^{-1}$
Mpc at the distance of A548.  The separation in velocity was less
clear: in fact the cluster A548 is reported to have a mean velocity of
$v=12394$ km/s (determined on 133 redshifts, Davis et al.  1995),
while A3367 had reported a value of $v=12780$ km/s (based on 6
velocities, Postman, Huchra \& Geller 1992). However, Postman \& Lauer
(1995) reported a velocity of $13461$ km/s for the brightest member of
A3367, clearly inconsistent with the above mean value. 

The cluster A548 [$\alpha(2000)=05^h 47^m 00^s$; $\delta(2000)=-25^o
36' 00''$] is a cluster of richness class 1 and Bautz-Morgan type
III. This cluster has been extensively studied both in the optical and
X--ray wavelength. Davis et al. (1995) reported a global velocity
dispersion of $903$ km/s, but this cluster appears dynamically very
complex. From the analysis of a mosaic of ROSAT PSPC observations,
Davis et al. (1995) found the presence of three extended sources
(dubbed S1, S2 and S3) with luminosities in the range $1.26-2.67\times
10^{43}$ erg/s in the $[0.1-2.4]$ keV band.  Performing a substructure
analysis of the optical sample, they detected three subcondensations:
two of these groups (labelled as $a$ and $b$ in their table 4b)
correspond to the extended X--ray emissions S1 and S2 respectively
(see their table 2).
These optical subclumps were already found by 
Escalera et al. (1994) with the use of a wavelet decomposition analysis.
They described A548 as a binary cluster, when large spatial scales 
are considered. Moreover, they detected at smaller scales a central subgroup.
These three components can be identified with the $a$, $c$ and $b$ 
substructures of Davis et al (1995), respectively. 
Another indication of the complex dynamical situation of this cluster 
appears from the different behaviour of galaxies with and without
emission lines in their spectra: Biviano et al. (1997) in their analysis of the
ENACS survey (Katgert et al. 1996, 1998) found a significant
offset between the mean velocities of these two types of objects.

On the contrary, the cluster A3367 was little studied so far. It has
coordinates $\alpha(2000)=05^h 49^m 40^s$; $\delta(2000)=-24^o 28'
00''$, is of richness class 0 and is classified as a Bautz-Morgan type
I-II. No other relevant date was found in the literature apart the
mean velocity reported above.

For these reasons we decided to concentrate our redshift survey on A3367
and on the region between A3367 and A548, and in this paper we present
a sample of 180 new radial velocities. 

The paper is organized as follows: in Sect. 2 we present the sample and the 
data reduction, in Sect. 3 we discuss the dynamical properties of the three 
peaks found in the sample and finally in Sect. 4 we summarize our results. 

 
\section{The sample}

\subsection{The photometric catalogue}

The starting photometric catalogue is the COSMOS/UKST galaxy catalogue
of the southern sky (Yentis et al. 1992), obtained from automated
scans of UKSTJ plates by the COSMOS machine. We extracted a circular
region of $2^o$ diameter, centered on $\alpha(2000)=05^h 48^m 34^s$
and $\delta(2000)=-25^o 17' 27''$, containing $11525$ objects to the
limiting magnitude $b_J<21.5$.

\begin{figure}[ht]
\psfig{figure=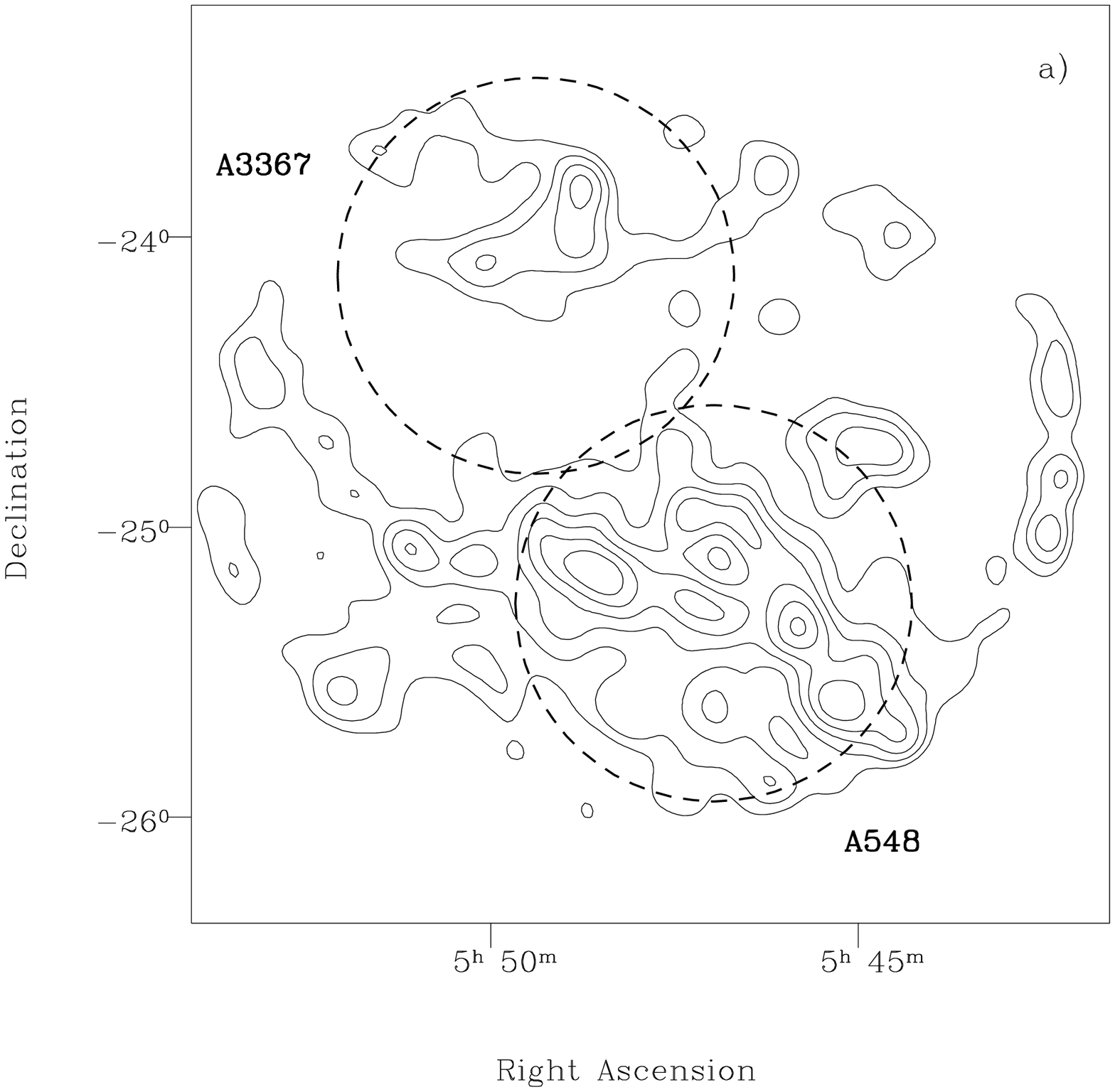,width=7.5truecm}
\psfig{figure=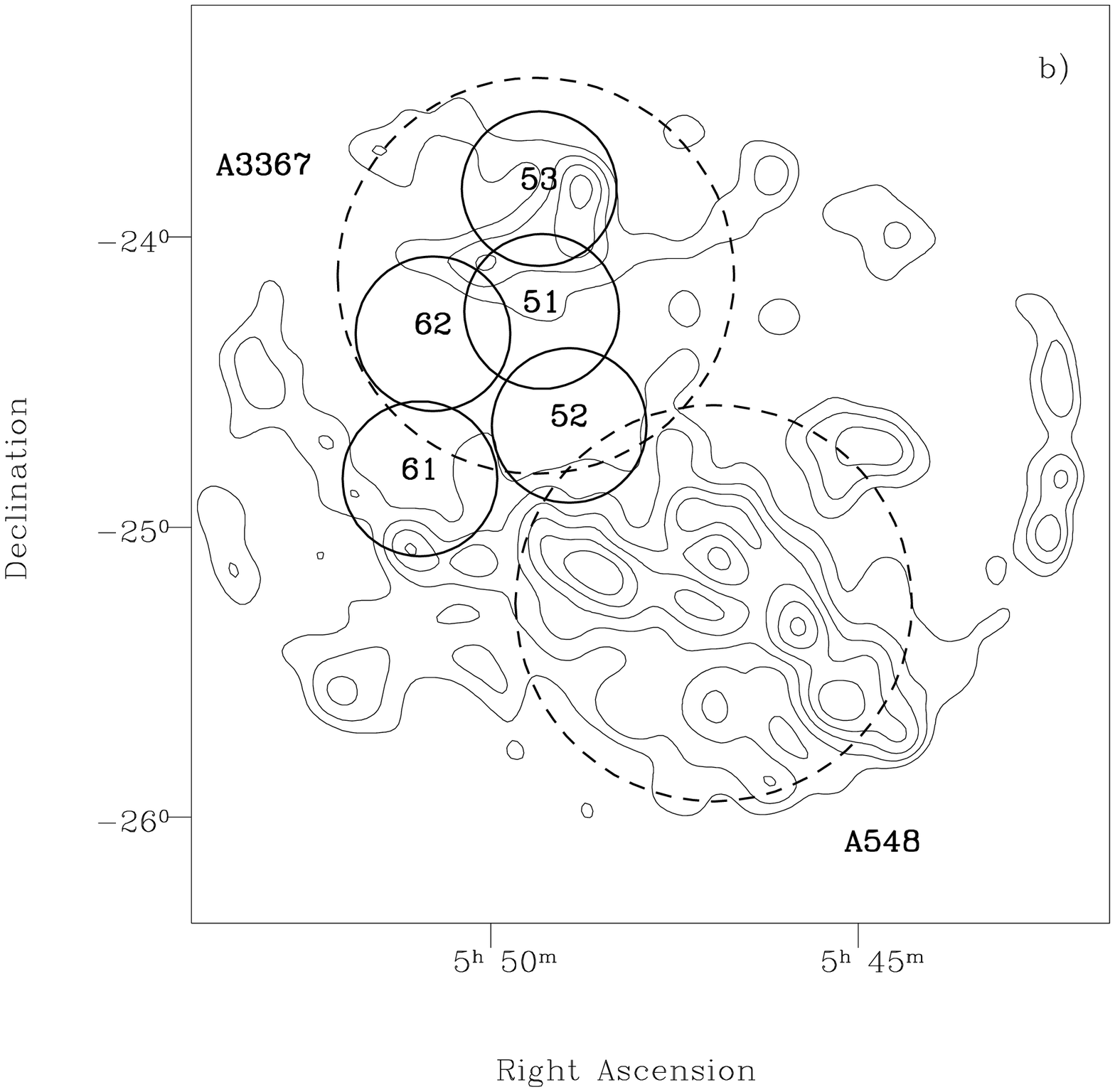,width=7.5truecm}
 	
\parbox[b]{7.5cm}{\caption {\label{fig:iso} {a) Isodensity contours of the 
region between A548 and A3367, centered on 
$\alpha(2000)=05^h 48^m 34^s$ and $\delta(2000)=-25^o 17' 27''$. 
The data are binned in 2 $\times$ 2 arcmin cells and smoothed with a Gaussian 
of 6 arcmin of FWHM. For the two clusters, dashed circles of 1 Abell radius 
have been drawn. 
b) As panel a), with the five OPTOPUS fields superimposed.}}}
  		 
\end{figure}

Fig. \ref{fig:iso}a shows the isodensity contours obtained binning the data in 
2 $\times$ 2 arcmin cells and smoothing with a Gaussian of 6 arcmin of FWHM.
For the two clusters circles of one Abell radius have been drawn around their 
nominal center. 
Note that it is already evident that A548 is not a smooth cluster with 
a single central nucleus, but presents multiple condensations.
Inside the Abell circle of A3367 we note a single condensation shifted 
northward with respect to the nominal center.
In Fig. \ref{fig:iso}b the same isodensity contours are shown with superimposed 
the OPTOPUS fields positions.

The coordinates of the centers of these fields are listed in columns (2) and 
(3) of Tab. \ref{tab:tvl1}, together with the observation date in column (4).

\begin{table}
\caption[]{\label{tab:tvl1} Observed OPTOPUS fields}
\scriptsize
\begin{flushleft}
\begin{tabular}{llll} 
\hline\noalign{\smallskip}
FIELD & $\alpha(2000)$ &$\delta(2000)$ & Date\\ \ \cr
\noalign{\smallskip} \hline\noalign{\smallskip}
 f51a & $05^h49^m24^s$     & $-24^o35'00''$    &  25/02/93\\
 f51b & $05^h49^m24^s$     & $-24^o35'00''$    &  26/02/93\\
 f52  & $05^h49^m00^s$     & $-25^o00'00''$    &  25/02/93\\
 f53  & $05^h49^m24^s$     & $-24^o10'00''$    & 26/02/93\\
 f61  & $05^h50^m46^s$     & $-25^o09'47''$    &  16/10/93\\
 f62  & $05^h50^m46^s$     & $-24^o39'52''$    & 17/10/93\\ 
\noalign{\smallskip}
\hline
\end{tabular}
\end{flushleft}
\end{table}

\subsection{Observations}

Spectroscopic measurements were obtained using the ESO 3.6m telescope at La 
Silla, equipped with the OPTOPUS multifiber spectrograph (Lund 1986), 
on the nights of 1993 February 25-26 and October 16-17.

The OPTOPUS multifiber spectrograph is formed by a bundle of 50 optical 
fibres at the Cassegrain focal plane of the telescope; this field has a 
diameter of 32 arcmin, and each fibre has a projected size on the sky of 2.5 
arcsec. We used the ESO grating $\#$15 with 300 lines/mm and a blaze 
angle of 4$^o$18$'$. This grating allows a dispersion of 174 \AA/mm in our
wavelength range  (3700--6100) \AA. We used the detector Tektronic 512 
$\times$ 512 CCD with a pixel size of 27 $\mu$m, corresponding to 4.5 \AA,
i.e. a velocity bin of $\simeq$ 270 km/s at 5000 \AA; the resolution is
$\sim 12$ \AA. 
Four of the 50 fibres were dedicated to sky measurements, leaving 46 fibres
available for the objects. 

Fields f51a, f51b, f52 and f53 were observed for 1 hour, split 
into two half-hour exposures in order minimize the presence of cosmic 
hits. Fields f61 and f62 were observed only for 1/2 hour.
The observing sequence was a 30-s exposure of a quartz halogen white lamp, a 
60-s exposure of helium vapour lamp for fields f61 and f62, 
or 60-s exposure of helium + neon vapour lamp for fields f51, f52 and f53; then
the scientific field, and again the arc and the white lamp.

\subsection{Data Reduction}

The extraction of the one-dimensional spectra was performed using the APEXTRACT 
package as implemented in IRAF\footnote{IRAF is distributed by the 
National Optical Astronomy Observatories, which is operated by AURA Inc. 
for the NSF.}.

Positions and tracing solutions of lamps and objects were determined on the 
flat field exposures.
The procedure we adopted to estimate the relative transmission of each fibre is 
based on the fitting of a Gaussian profile to the [OI]$\lambda$5577 sky line in 
each spectrum and on computing the continuum-subtracted
flux of this line (Bardelli et al. 1994).
If we assume that the flux and the shape of the spectrum of the night sky remain
constant in the telescope field, this value is the same in each spectrum apart 
from the transmission of the fiber, which is a multiplicative factor. 
After having normalized the spectra, we can subtract the `mean sky' obtained as
the average of the 4 sky spectra.

\subsection{Redshift data} 

We have obtained a total of 276 spectra: 45 were not useful for redshift
determination ($16\%$ of the total), because of poor signal--to--noise ratio 
or badly connected fibers, and 51 turned out to be stars ($22\% $ of the
reliable spectra), leaving us with 180 galaxy redshifts. The galaxies whose
spectrum presents detectable emission lines are 79, corresponding to a
percentage of $44\%$ of the total.

The radial velocities of galaxies with spectra with absorption
lines have been determined using the program XCSAO in the IRAF task RVSAO
(Kurtz et al. 1992), which is based on the cross-correlation method of Tonry
\& Davis (1979).
The determination of redshift is done by fitting a parabola to the main peak of 
the cross-correlation function.
Sixteen different templates (eight stars and eight galaxies) were
used for the determination of the radial velocities, choosing as better 
estimate the one which gave the minimum cross-correlation error, defined as:

\begin{equation}
\epsilon=\frac{\displaystyle 3}
           {\displaystyle 8}\frac{\displaystyle w}{\displaystyle (1+r)}
\end{equation} 
where $w$ is the FWHM of the cross-correlation peak and $r$ is the ratio 
between the height of the correlation peak and the $rms$ of the antisymmetric 
part of the correlation function (Kurtz et al. 1992). 

To estimate the redshift of spectra with strong emission lines we used the 
EMSAO program in the IRAF task RVSAO.

\begin{figure}[ht]
\psfig{figure=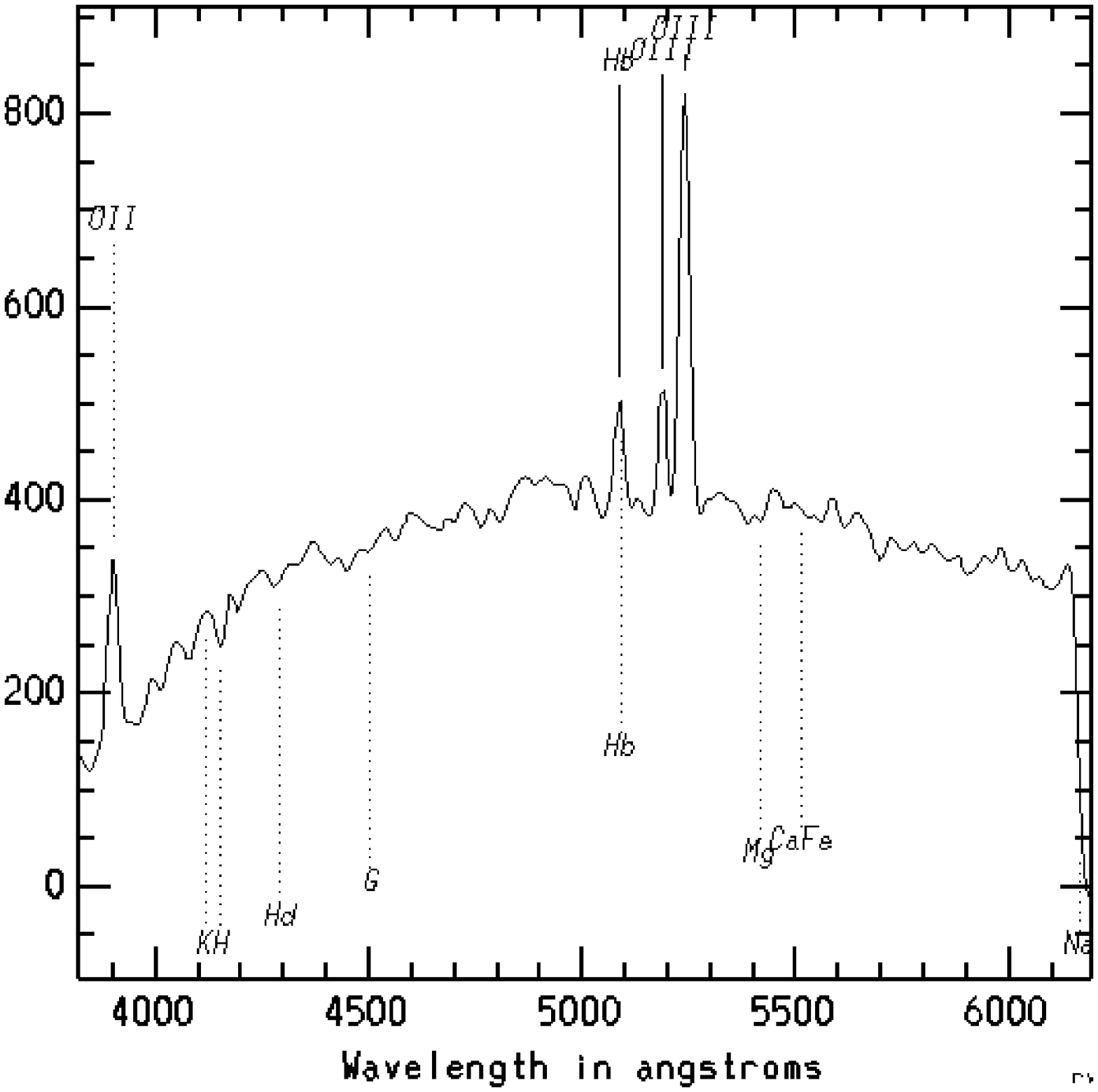,angle=90,width=7.truecm,bbllx=38pt,bblly=125pt,bburx=574pt,bbury=666pt}
\psfig{figure=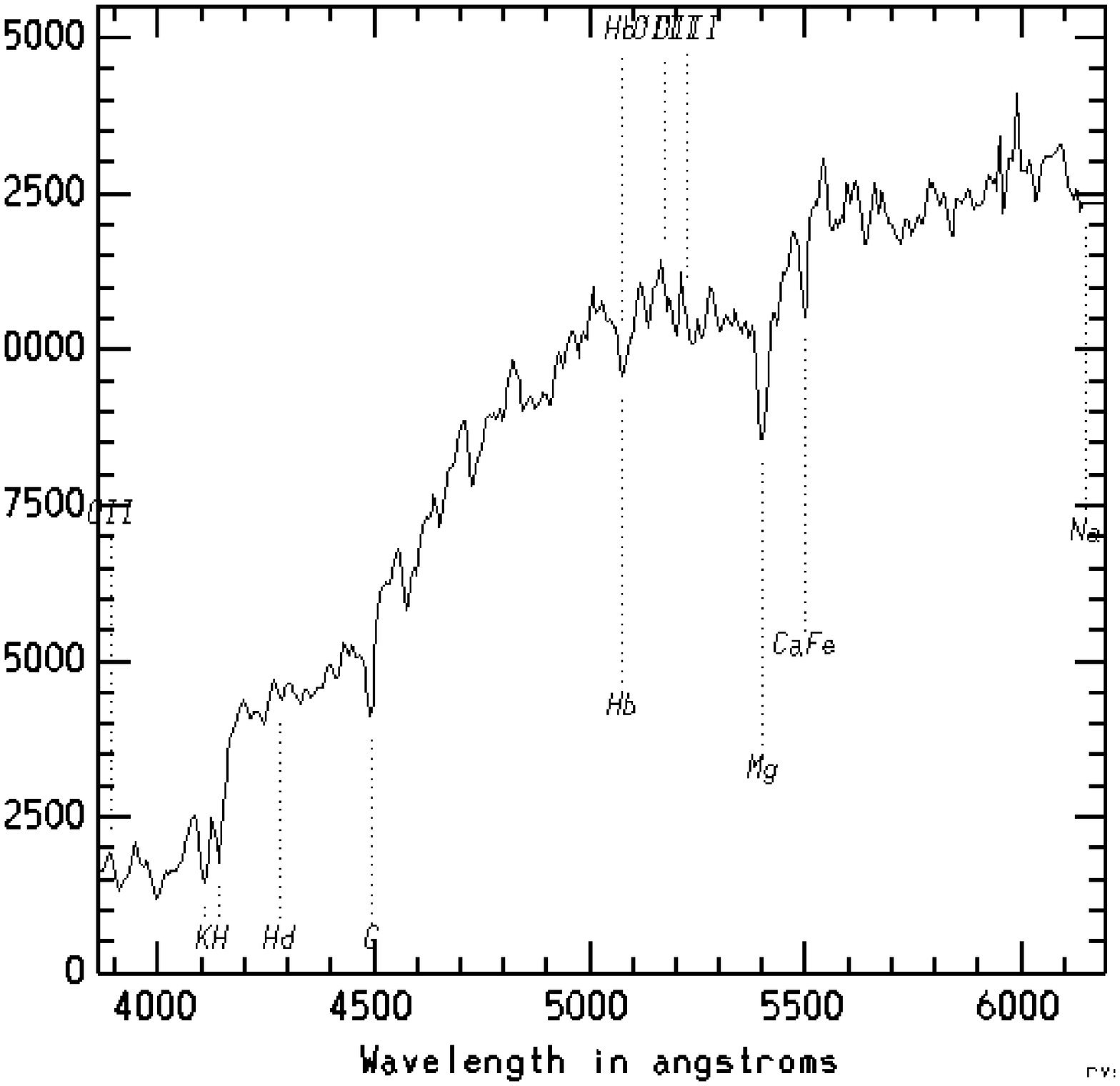,angle=90,width=7.truecm,bbllx=38pt,bblly=122pt,bburx=575pt,bbury=670pt} 	
\parbox[b]{7.5cm}{\caption {\label{fig:spe} {Top: Example of a spectrum with 
strong emission lines ($v=13954\pm$29 km/s). Bottom: Example of a spectrum with 
only features in absorption ($v=13106\pm$25 km/s, $r=17.5$). The spectra
are plotted in arbitrary units on y-axis}}}
\end{figure}

The top panel of Fig. \ref{fig:spe} shows an example of a spectrum with strong 
emission features, with [OII]$\lambda 3727$, [H$\beta$]$\lambda 4861$,
[OIII]$\lambda 4959$, $\lambda5007$ lines, while in the bottom panel of
Fig. \ref{fig:spe} a spectrum with only absorption lines is presented.

In Tab. \ref{tab:511} - \ref{tab:62} we list the galaxies with redshift
determination. 
Columns (1), (2) and (3) list the right ascension, the declination and the 
$b_J$ magnitude respectively; column (4) and (5) give the heliocentric velocity 
($v=cz$) and its internal error (in km/s), from absorption and emission 
lines respectively. The code in column (6) indicates the presence
of emission lines: the symbols a, b, c, d, refer to [OII]$\lambda$3727\AA, 
[H$\beta$]$\lambda$4861\AA, [OIII]$\lambda$4959\AA\  and
[OIII]$\lambda$5007\AA, respectively. 
 
\begin{table}
\caption[]{\label{tab:511}Field 51a}
\scriptsize
\begin{flushleft}
\begin{tabular}{ccrllc} 
\hline\noalign{\smallskip}
$\alpha$(2000) & $\delta$(2000) & $b_J$  &$v_{abs}$ km/s & 
$v_{emis}$ km/s & notes \\ 
\noalign{\smallskip}
 \hline\noalign{\smallskip}
05 49 54.0 & -24 23 19.9  & 15.8 & 13106$\pm$25 &            &       \\
05 49 27.0& -24 46 14.3  & 16.0 & 13367$\pm$48 &             &       \\
05 48 22.7  & -24 33 15.1  & 18.2 & 30431$\pm$85 &  &   \\ 
05 49 22.2  & -24 35 25.2  & 18.3 & 30771$\pm$38 & &   \\ 
05 49 42.6  & -24 41 47.2  & 17.5 & 13784$\pm$48 &            &       \\
05 49 41.1  & -24 34 31.0 & 18.3 & 19647$\pm$25 &     	    &       \\
05 50 06.4  & -24 23 40.7  & 17.9 &             & 13803$\pm$50  &  abcd  \\ 
05 49 34.6  & -24 30 57.2  & 16.6 & 19632$\pm$22 &             &       \\       
05 49 07.6  & -24 34 27.5  & 18.3 & 30596$\pm$26 &            &       \\
05 49 39.7  & -24 25 26.3  & 16.0 &             & 12237$\pm$53&   abcd \\ 
05 49 59.8  & -24 28 05.9  &19.0 &             & 30797$\pm$33  &   abcd \\ 
05 49 41.3  & -24 32 42.9  &15.7  & 13363$\pm$29 &              &         \\
05 48 26.6  & -24 42 09.9  & 17.3 & 13363$\pm$20 &              &         \\  
05 49 36.3  & -24 30 18.9  & 18.7 & 19837$\pm$49 &               &        \\
05 48 25.5  & -24 41 43.9  & 18.8 &           & 13954$\pm$29   &  abcd \\ 
05 49 50.8  & -24 39 55.1  &16.9  &  13289$\pm$32&              &     \\
05 48 48.6  & -24 22 42.6  & 14.9 &  11895$\pm$51 &           &        \\
05 49 10.2  & -24 38 22.8  &18.6  &  42066$\pm$63 &           &   \\ 
05 48 23.1  & -24 32 32.8  & 17.7 &  30261$\pm$70 &            &   \\
05 49 37.5  & -24 32 10.5  & 18.5 &           & 42783$\pm$179   &     abcd\\ 
05 48 25.5  & -24 43 15.1  & 18.1 &          & 13950$\pm$22  &   abcd\\ 
05 50 17.9  & -24 28 38.1  & 17.5 &           & 11891$\pm$36  &       ad\\ 
05 49 37.4  & -24 33 33.2  & 17.9 &  19913$\pm$21 &             &      \\
05 48 50.2  & -24 21 06.9  & 16.2	 &  13741$\pm$30  &     &  \\ 
05 49 39.5  & -24 39 08.2  & 18.8	 &  39649$\pm$105 &      & \\
05 49 26.1  & -24 22 25.5  &17.4	 &  19498$\pm$20  &     &      \\
05 48 37.0 & -24 26 36.8  &18.6 &  42688$\pm$62   &      &  \\
05 49 14.8  & -24 25 52.1  & 17.8	 &        & 13447$\pm$59  &  bcd \\
05 48 22.1  & -24 41 43.7  & 16.8	 &  13706$\pm$77  &       &   \\
05 50 12.8  & -24 40 08.6 & 18.9	 &            & 41774$\pm$15 &  abcd \\ 
05 49 41.2  & -24 34 06.4  &18.1	 &  11803$\pm$60  &     &  \\
05 48 48.7  & -24 29 46.0 &18.6 &  30758$\pm$28   &      & \\
05 49 53.8  & -24 27 58.7  &16.6	 &  13923$\pm$110   &        & \\
05 49 20.2  & -24 38 39.2  &18.6	 &  13684$\pm$136    &     &  \\ 
05 48 42.0 & -24 26 50.3  &18.7 &                & 12119$\pm$69    &  abd\\ 
05 50 05.1  & -24 28 47.5  &16.5	 &  12131$\pm$50    &       & \\
05 49 12.2  & -24 29 59.5   &18.7	 &  51690$\pm$52    &       & \\
05 49 04.5  & -24 29 01.0  &17.5	 &  13529$\pm$112    &     & \\
05 50 03.6  & -24 38 14.1   &15.0	 &  13368$\pm$26    &      & \\
\noalign{\smallskip}
\hline
\end{tabular}
\end{flushleft}
\end{table}

\begin{table}
\caption[]{\label{tab:512}Field 51b}
\scriptsize
\begin{flushleft}
\begin{tabular}{ccrllc} 
\hline\noalign{\smallskip}
$\alpha$(2000) & $\delta$(2000) & $b_J$  & $v_{abs}$ km/s & 
$v_{emis}$ km/s & notes \\ 
\noalign{\smallskip}
 \hline\noalign{\smallskip} 
05 49 08.5  & -24 36 49.2  & 20.0   &   53764$\pm$36 &     &\\
05 49 16.7  & -24 38 47.7  & 20.1	&   42217$\pm$36 &   &\\
05 48 24.8  & -24 34 23.8  & 20.1	&   30341$\pm$55 &  	& \\
05 50 03.6  & -24 48 00.9  & 20.1	&  & 41842$\pm$69 &   acd \\
05 49 08.4  & -24 44 27.7  &19.6	&  & 30031$\pm$27  &   acd \\ 
05 50 16.0 & -24 26 50.6  & 20.1	&  & 53060$\pm$80   &  a \\ 
05 49 03.4  & -24 19 56.0 & 19.8	&   &  80128$\pm$80 &	    a \\ 
05 50 33.7  & -24 35 35.7  & 19.6	&  &  41023$\pm$77 &  	 abcd \\
05 48 47.0 & -24 24 14.5  & 19.9	& &   40152$\pm$50 &  ab \\ 
05 49 48.0 & -24 26 13.0 & 20.1&  & 17267$\pm$85 &  c \\
05 49 29.1  & -24 21 22.5  & 20.1   &  19508$\pm$69 &	         &     \\
05 50 01.6  & -24 37 33.2  & 19.7	&   &  13274$\pm$77 &	 abcd \\ 
05 48 45.4  & -24 22 12.6  & 19.9	&   &  66537$\pm$80 &	     a \\ 
05 49 58.5  & -24 39 02.4  & 20.0	&   & 19932$\pm$17 & acd \\ 
05 48 56.5  & -24 26 15.2  & 19.8	&  &  88903$\pm$80 &  a \\ 
05 49 30.9  & -24 33 17.8  & 20.2	&   & 65483$\pm$28 & abcd\\ 
05 48 36.6  & -24 46 25.4  & 19.8	& 50943$\pm$120	&   & \\  
05 50 33.0 & -24 34 18.7  & 20.0	&  & 34930$\pm$22 &  ab\\ 
05 48 38.1  & -24 25 11.6  &19.8	& 31182$\pm$72	&  &\\ 
05 50 19.4  & -24 38 06.4  & 19.9  & 86756$\pm$84  &  & \\
05 49 58.0 & -24 29 21.7  & 20.0 & 81631$\pm$75 & 81612$\pm$80 &   a \\
05 48 51.7  & -24 21 40.5  & 19.7  & 66526$\pm$91	&  66285$\pm$80&   a \\
05 48 49.4  & -24 21 49.2  & 19.7  &   & 13598$\pm$56 &   abcd\\ 
05 49 42.0 & -24 21 14.7  & 20.0 & 66954$\pm$52	 &  \\
05 49 39.0 & -24 23 46.6  & 19.8  &  51356$\pm$62  &	   &\\
\noalign{\smallskip}
\hline
\end{tabular}
\end{flushleft}
\end{table} 
\begin{table}
\caption[]{\label{tab:52}Field 52}
\scriptsize
\begin{flushleft}
\begin{tabular}{ccrllc} 
\hline\noalign{\smallskip}
$\alpha$(2000) & $\delta$(2000) & $b_J$  & $v_{abs}$ km/s & 
$v_{emis}$ km/s & notes \\ 
\noalign{\smallskip}
 \hline\noalign{\smallskip} 
05 49 10.9  & -24 53 17.3  & 17.5	&   19748$\pm$47  &  \\
05 49 08.9  & -25 13 50.5  & 19.8	&  13679$\pm$106  &  \\
05 49 36.2  & -24 51 09.5  & 17.3   &   &    13394$\pm$51 &	   abcd \\ 
05 48 26.3  & -25 12 47.1  & 17.2	&  11039$\pm$105 &  &  \\
05 48 48.2  & -25 11 48.8  & 18.9	&    &   11949$\pm$84	&  acd \\
05 48 36.7  & -24 54 18.4  & 19.1	&  50873$\pm$86	&  \\
05 49 24.5  & -25 02 09.4  & 19.1   &    & 30093$\pm$58 &  	acd  \\ 
05 49 33.7  & -25 07 09.6  & 18.3   &  16959$\pm$81	& 16890$\pm$49 	& acd\\ 
05 49 54.5  & -25 03 54.6  & 18.9	&  12048$\pm$131 &   & \\
05 49 29.6  & -25 00 53.2 & 18.0	&  13255$\pm$120 & & \\
05 49 55.1  & -24 56 55.4  & 18.0	&  13449$\pm$93	&  &\\
05 49 34.4  & -25 04 33.3  & 16.9	&  14350$\pm$33	&  &\\
05 49 15.3  & -25 07 57.7  & 19.2	& 11727$\pm$104	&  &\\
05 48 36.3  & -25 01 19.1  & 19.7	&  50542$\pm$67	&  &\\
05 48 37.5  & -24 53 33.6  & 18.8	&   & 30361$\pm$26 &  abcd \\ 
05 48 35.9  & -25 10 09.7  & 18.8	&  22404$\pm$35	&  &\\
05 48 22.7  & -25 12 17.0 & 18.1	&  13187$\pm$25	&  &\\
05 49 22.9  & -24 58 50.4  & 19.5	&    & 12721$\pm$41 &  	ad\\ 
05 48 41.1  & -25 09 13.4  & 18.0   & 12332$\pm$91	&  &\\
05 48 58.5  & -25 06 08.5  & 19.7	 &    & 22400$\pm$73	&  	abcd\\ 
05 50 01.2  & -24 53 29.7  & 17.6	 &  13521$\pm$93 &  \\
05 48 04.8  & -24 57 34.5  & 19.4	 &    & 52703$\pm$124	&  abcd \\
05 49 09.5  & -25 03 22.5  & 19.3	 &    & 12213$\pm$54	& abd \\ 
05 48 30.0 & -24 50 20.6  & 18.9	 &   & 11886$\pm$100	&  ac \\ 
05 48 14.5  & -25 07 54.8  & 19.6	  &   &  37211$\pm$150 &   ad \\
05 48 58.5  & -24 49 56.1  & 19.4	  &   &  29678$\pm$129	& abcd \\ 
05 48 47.0 & -25 09 16.8  & 18.6	  &    & 13229$\pm$66	&    ac \\
05 49 40.6  & -25 06 27.5  & 18.3	  & 48574$\pm$72 & 	 &      \\
05 49 18.6  & -25 15 22.1  & 19.6	  & 12320$\pm$95&	 \\
05 49 23.6  & -25 03 36.1  & 19.7	  &   &  27429$\pm$58 &	 ac\\ 
05 48 40.0 & -24 59 20.0 & 17.0 &    13523$\pm$55 &	  &\\
05 49 40.8  & -25 10 56.0 & 19.8	  &   &   114834$\pm$80 &   a \\
05 48 30.3  & -24 52 08.8  & 19.8	  &    &  13229$\pm$58 & abcd\\ 
05 48 02.5  & -25 01 20.5  & 16.2	  &    & 12018$\pm$131 & ac \\
\noalign{\smallskip}
\hline
\end{tabular}     
\end{flushleft}
\end{table}

\begin{table}
\caption[]{\label{tab:53}Field 53}
\scriptsize
\begin{flushleft}
\begin{tabular}{ccrllc} 
\hline\noalign{\smallskip}
$\alpha$(2000) & $\delta$(2000) & $b_J$  & $v_{abs}$ km/s & 
$v_{emis}$ km/s & notes \\ 
\noalign{\smallskip}
 \hline\noalign{\smallskip} 
05 49 05.5  & -24 06 43.5   & 19.3 &  31145$\pm$31	&  & \\
05 48 49.2  & -24 15 18.8   & 17.6	 &  29392$\pm$39  &   &\\
05 50 30.9  & -24 05 50.7   & 19.1	 &  30655$\pm$30 &  &\\
05 48 51.6  & -24 06 45.8   & 19.3	 & 30283$\pm$36	 & &\\
05 48 55.1  & -24 13 17.0  & 18.8	 & 31518$\pm$31	&  &\\
05 49 15.7  & -24 01 41.6   & 18.9	  &  &  13810$\pm$56	& ad\\ 
05 49 20.8  & -24 03 49.8   & 17.1	  & 30057$\pm$61	&  &\\
05 48 53.5  & -24 02 07.2   & 19.4	  & 30506$\pm$96	&  &\\
05 48 47.5  & -24 10 52.4   & 20.6	 & 30588$\pm$36	&  &\\
05 48 49.4  & -24 09 12.6   & 19.4	 &  31078$\pm$42	&  &\\
05 49 26.7  & -24 17 06.9   & 17.8	  & 29254$\pm$79	 & &\\
05 49 04.8  & -24 04 29.3   & 16.0	  & 13618$\pm$59&	  &\\
05 49 55.3  & -24 05 31.2   & 18.1	  &   & 30080$\pm$73 &	  abcd \\ 
05 49 17.3  & -24 18 31.9   & 19.3	  &    & 77349$\pm$27	&  ab\\ 
05 48 45.9  & -24 13 33.4   & 19.1	  & 31014$\pm$49 &	  &\\
05 50 03.0  & -24 01 20.0  & 16.5	  & 8981$\pm$84 &  8917$\pm$15& abcd\\ 
05 50 20.7  & -24 16 15.4   & 19.4	  & 42073$\pm$63  &	 &\\
05 49 55.3  & -24 08 03.3   & 19.2	  &   & 30121$\pm$23  &	  abcd \\ 
05 48 38.1  & -24 20 42.8   & 19.4	  & 31206$\pm$29 &	  &\\
05 49 13.4  & -24 02 46.0  & 19.3	  & 93404$\pm$60  &	 &\\
05 48 50.8  & -24 20 21.1   & 16.5	  & 19944$\pm$25  &	  &\\
05 48 58.1  & -24 05 06.7   & 18.9	 &  30472$\pm$28 &	  &\\
05 48 57.5  & -24 17 25.1   & 18.1	  & 31429$\pm$43 &	  &\\ 
05 50 29.7  & -24 12 45.1   & 19.1	  &   &  13870$\pm$37  &abcd\\ 
05 48 41.2  & -24 15 01.2   & 18.1	  & 30736$\pm$32  &	  &\\
05 48 36.8  & -24 07 04.5   & 19.5	  & 31040$\pm$71 &	&\\
05 49 02.0 & -24 18 06.2   & 17.1	  &  &  13120$\pm$65  &	  abd \\ 
05 49 17.8  & -23 56 52.7   & 19.5	 &  29940$\pm$39  &	  &\\
05 49 07.7  & -24 09 30.9   & 18.9	 &  3024$\pm$238  &	  &\\
05 48 57.0 & -24 09 53.4   & 17.5	  & 29551$\pm$112 & 29345$\pm$111&ad\\ 
05 49 25.1  & -24 03 23.5   & 18.7	  & 42086$\pm$69 & 42136$\pm$80&    a\\ 
05 49 43.9  & -24 10 55.1   & 18.3	  & 31847$\pm$110 &   &    \\
05 49 01.7  & -24 04 44.3   & 18.3	  & 30776$\pm$33	&   &\\
\noalign{\smallskip}
\hline
\end{tabular}     
\end{flushleft}
\end{table}

\begin{table}
\caption[]{\label{tab:61}Field 61}
\scriptsize
\begin{flushleft}
\begin{tabular}{ccrllc} 
\hline\noalign{\smallskip}
$\alpha$(2000) & $\delta$(2000) & $b_J$  &$v_{abs}$ km/s & 
$v_{emis}$ km/s & notes \\ 
\noalign{\smallskip}
 \hline\noalign{\smallskip} 
05 51 23.0 & -25 23 13.8  & 15.9 & 42806$\pm$44	 & &\\
05 50 15.9  & -25 16 15.4  & 19.0 & 11696$\pm$68	 &    &   \\  
05 51 28.1  & -25 24 16.9  & 18.3 & 40426$\pm$92	 & 40383$\pm$80 &   a\\ 
05 51 23.7  & -25 24 59.0 & 19.9 & 41583$\pm$98	 &  &\\
05 50 22.3  & -25 18 09.2  & 17.2 & 11595$\pm$28	&  &\\
05 51 18.0 & -25 15 40.8  & 17.9 &    &  12124$\pm$40	&  abcd\\ 
05 51 55.3  & -25 14 14.2  & 19.6	  &   &  44811$\pm$20	&  ad \\ 
05 50 01.8  & -25 18 49.2  & 19.1	  &    & 11779$\pm$62	& acd \\ 
05 51 07.9  & -25 09 31.7  & 17.9	    &  & 11769$\pm$73	 &  abcd \\ 
05 52 04.2  & -25 12 22.5  & 19.7	  & 45116$\pm$112	 & &\\
05 51 52.9  & -25 12 16.1  & 19.2	   &  45788$\pm$66	 & &\\
05 50 13.7  & -25 20 11.6  & 19.6	  & 30356$\pm$58    &  &\\
05 50 57.1  & -25 01 13.6  & 17.0	  & 26279$\pm$53& 26228$\pm$35 & bcd \\ 
05 51 16.6  & -25 19 45.4  & 19.6	   &  42172$\pm$61	&  &\\
05 50 17.0 & -24 58 37.4  & 19.9	   &    &  17715$\pm$106 & ad\\ 
05 50 14.2  & -25 13 55.0 & 18.2	  & 11739$\pm$69	& &\\
05 51 13.2  & -25 24 22.3  & 18.7	  &   & 11652$\pm$45	&  abcd \\ 
05 50 47.1  & -25 04 50.3  & 20.6	  & 26277$\pm$82	&  &\\
05 50 41.0 & -25 03 56.6  & 19.3	 &    &  59525$\pm$80	&a \\ 
05 51 14.9  & -25 24 57.3   & 19.6	 &    &  17674$\pm$123	& abcd\\ 
05 50 15.6  & -25 10 02.8 & 18.7	&  & 29035$\pm$34  &   abcd \\ 
05 50 42.0 & -25 20 04.4     & 19.7 & 61220$\pm$56	&  &\\
05 51 19.9  & -25 12 19.8     & 19.1 &    & 44958$\pm$80	& a\\
05 51 46.0 & -25 18 32.5    & 17.7	  & 13427$\pm$87 & 13428$\pm$86&  a\\ 
05 50 31.0 & -25 01 34.4    & 18.5	  &   &  11811$\pm$52	 & ad\\ 
05 51 28.4  & -25 18 49.0   & 19.8	  &    & 46585$\pm$96	&     ab \\ 
05 49 59.7  & -25 12 48.1   & 19.3	  &    &  16948$\pm$68	&  abcd \\
05 50 17.3  & -25 02 05.4   & 18.3	  & 40337$\pm$35	&  &\\		
05 50 58.0 & -24 56 24.2    & 17.4	   &29049$\pm$51	& & \\
05 50 01.8  & -25 08 04.5    & 19.7	   &   & 12402$\pm$132	&	 a \\
05 52 03.1  & -25 03 22.1     & 17.2  & 41200$\pm$105	&  &\\
05 50 46.0 & -25 15 40.7     & 17.7	  & 11883$\pm$64	&  &\\
05 51 31.6  & -25 23 24.6     & 19.4	  & 41210$\pm$44	&  &\\
\noalign{\smallskip}
\hline
\end{tabular}     
\end{flushleft}
\end{table}

\begin{table}
\caption[]{\label{tab:62}Field 62}
\scriptsize
\begin{flushleft}
\begin{tabular}{ccrllc} 
\hline\noalign{\smallskip}
$\alpha$(2000) & $\delta$(2000) & $b_J$  &$v_{abs}$ km/s & 
$v_{emis}$ km/s & notes \\ 
\noalign{\smallskip}
 \hline\noalign{\smallskip} 
05 50 39.5  & -24 28 33.7    & 19.7	  & 52479$\pm$121 &	  & \\
05 50 56.6  & -24 34 45.0    & 17.8	  &              & 11417$\pm$39 &abcd\\ 
05 51 38.5  & -24 31 02.5    & 17.5       & 13219$\pm$28  &	& \\
05 51 01.0  & -24 38 24.4    & 18.9	  & &  13594$\pm$65&  ad  \\ 
05 51 20.3  & -24 40 20.0    & 20.0 & 53185$\pm$91  &	& \\
05 50 20.8  & -24 48 31.4    & 18.9	  & 48556$\pm$78  & 	  & \\
05 50 17.3  & -24 52 30.0    & 20.0	  &   & 19575$\pm$17  &	     abd \\ 
05 50 51.9  & -24 52 28.2    & 19.2	&   &19535$\pm$31 &  abcd \\
05 51 36.5  & -24 46 04.5    & 20.2	&   & 26141$\pm$90 & ad  \\ 
05 51 46.5  & -24 35 31.4    & 18.1	  & 28357$\pm$61 & 28398$\pm$76 &  a\\ 
05 51 02.5  & -24 46 20.4    & 18.8	  & 30549$\pm$47  &	 &\\
05 50 30.0  & -24 30 50.5    & 20.3	  &   &  40183$\pm$60 &	abcd \\
05 50 13.7  & -24 51 10.4    & 20.0 	&  & 69130$\pm$80  &	a    \\
05 51 01.4  & -24 49 48.0    & 20.3	  &    & 66474$\pm$80  &	a \\ 
05 51 03.1  & -24 54 17.2    & 18.7	  &   & 17724$\pm$80	&a \\ 
05 50 38.3  & -24 24 58.4    & 20.1	  &88022$\pm$98 &	&	  \\
\noalign{\smallskip}
\hline
\end{tabular}     
\end{flushleft}
\end{table}

We remember that the cross-correlation errors are only internal formal errors. 
In order to have true statistical errors, these values 
have to be multiplied for the factor 1.53 found by Vettolani et al. (1998)
comparing multiple observations of the same galaxies: 
after this correction, the average statistical error on our velocities 
is $\simeq$ 95 km/s. If one wants to take into account also the uncertainties 
introduced by the different reduction procedures, the factor is slightly 
larger and
has the value of $\sim 1.9$ (see Bardelli et al. 1994).

\begin{figure}[ht]
\psfig{figure=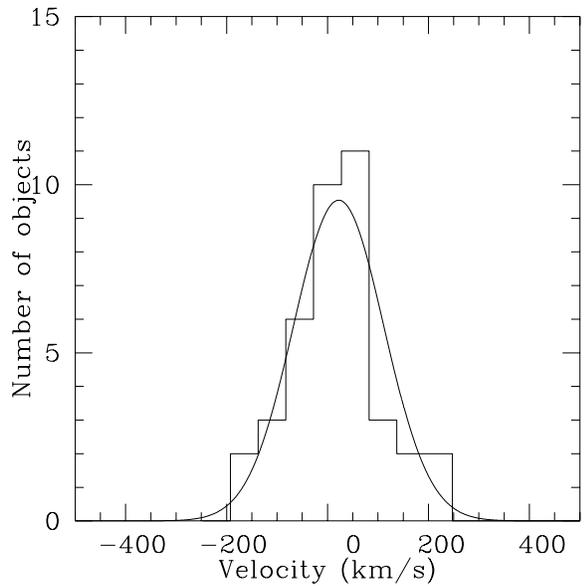,angle=0,width=8.truecm}
\parbox[b]{7.5cm}{\caption {\label{fig:star} {Histogram of the velocities of 
41 stars observed by chance in our survey. The superimposed solid curve 
corresponds to a Gaussian with $< v >$= 22 km/s and $\sigma$ = 90 km/s.}}}
\end{figure}

In order to check the zero point precision of our velocity scale, 
we considered the histogram of the measured velocities of the stars 
misclassified as galaxies (Fig. \ref{fig:star}), which are expected to
have a zero mean velocity.
Considering only the 41 spectra with the higher signal-to-noise
ratio, we found  $< v >_{stars} = 22 \pm 14$ km/s ($\sigma_{stars}=90$ km/s): 
this small systematic effect will be neglected in the following analysis, 
since the errors associated to the galaxy velocities are larger. 
However, we can not exclude that the value of $< v >_{stars}$ is completely 
due to bulk motions of stars in this region of the sky.

Very recently, Cappi et al. (1998), analysing the ESP survey (Vettolani et al.
1997, 1998), noted a 
systematic difference between the velocities estimated from the emission lines 
and the cross-correlation for the same galaxy, with an average difference 
of $<v_{abs}-v_{emiss}>=93 \pm 6$ km/s (obtained from more than 700 galaxies).  
Our observations are taken in the same instrumental configuration of the ESP
survey and can give an independent estimate of this effect,
although with a smaller sample. On the basis of 10 galaxies, we find 
$<v_{abs}-v_{emiss}>=60 \pm 30$ km/s, consistent within the errors
with the result of Cappi et al. (1998). 

\section{Discussion}

\begin{figure}[ht]
\psfig{figure=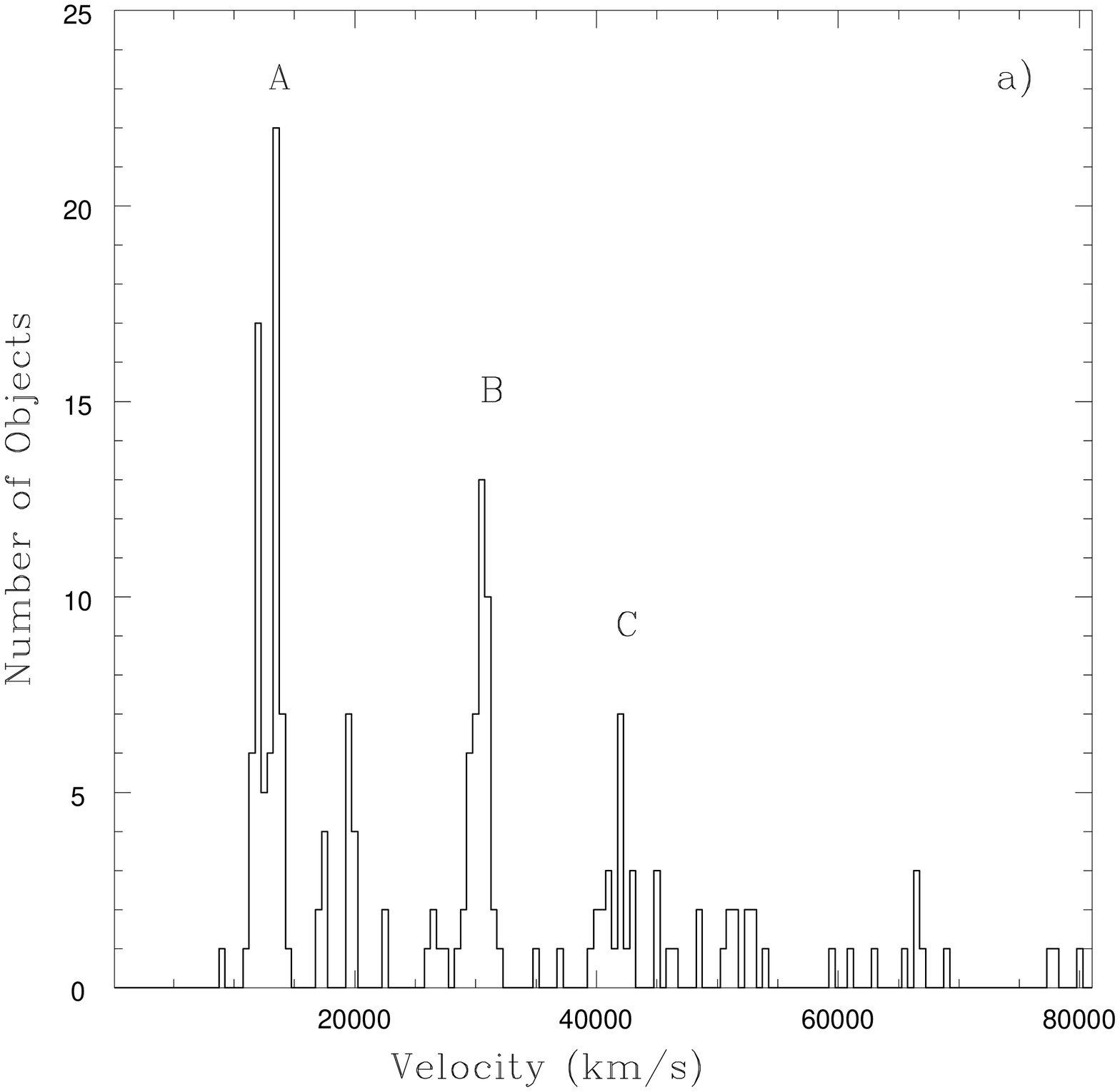,angle=0,width=8.truecm}
\psfig{figure=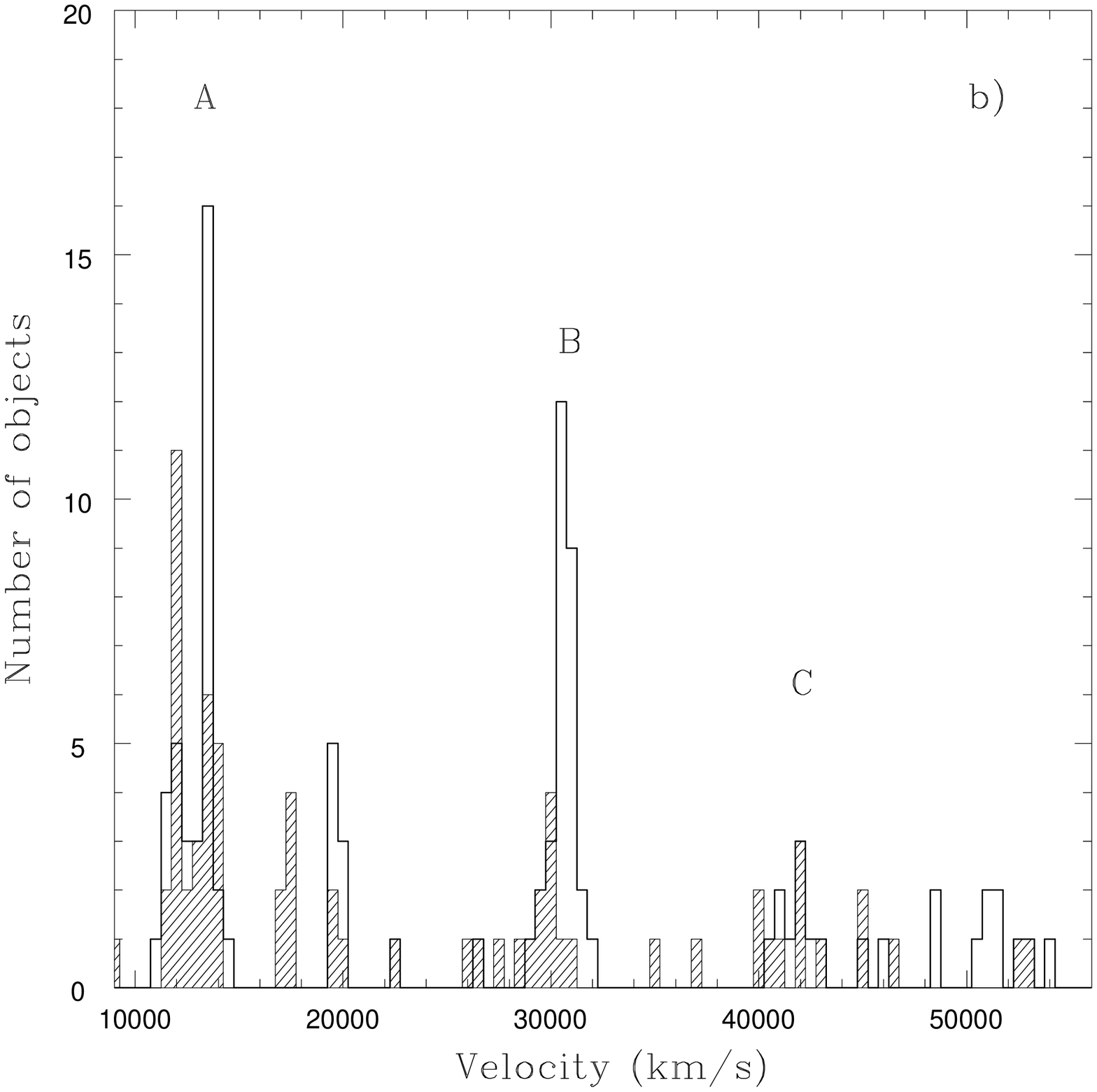,angle=0,width=8truecm}

\parbox[b]{7.5cm}{\caption {\label{fig:ist1}{a) Histogram of the observed 
velocities in bins of 500 km/s for all galaxies of our sample. b) Close up
of panel a); the shadowed histogram represents the distribution of galaxies
with emission lines, while the open histogram refers to the
galaxies with only absorption lines.}}}
\end{figure}

In Fig. \ref{fig:ist1}a the histogram of the galaxy velocities is shown.
It is clear the presence of at least three peaks (labelled as A, B, C in the
figure): the first is at a velocity 
of $\sim 13000$ km/s, the second and the third at $\sim 30000$ km/s and
$\sim 40000$ km/s, respectively. Peak A is at the same velocity of A548 and 
presents a clear bimodality. 

Although no significant differences (through a K-S test) are found between the
overall distributions of galaxies with and without emission lines, a more
detailed analysis of the three peaks reveals that the two distributions
inside the single peaks are in fact different (see Fig.\ref{fig:ist1}b).
In particular, for peak A it is evident a separation in velocity, being
the population of emission line objects dominant in the clump 
at lower velocity: the percentage of emission line galaxies with respect to
the total in this clump is $54\%$, while it is $39\% $ in the higher velocity 
clump. 
Because galaxies with and without emission lines have different luminosity
functions (Zucca et al. 1997), it could be suspected that their different 
distribution is a consequence of a change in the relative values of their 
selection functions: however, the width of peak A is relatively narrow (less 
than $1000$ km/s) and therefore this effect is more likely due to a real 
variation in morphological composition in the two subclumps.

We estimated the dynamical parameters (mean velocity and velocity dispersion) 
of the three peaks with the biweight estimators of location and scale 
(Beers et al. 1990). The advantage of these 
estimators, with respect to the standard mean and dispersion, is that of
minimizing biases from interlopers, giving less weight to data with higher
distance from the median. The confidence intervals of the two estimators 
are calculated boostrapping the data with 100 random catalogs. 
In order to find the velocity range in which the cluster members lie, we have 
assumed that the velocity distribution of cluster galaxies is Gaussian, as 
expected when the system has undergone a violent relaxation (see details in 
Bardelli et al. 1994). 

For the case of peak A, in which the presence of a substructure was suspected
on the basis of both a visual inspection of the velocity histogram and
the shape estimators $a$, $b_1$, $b_2$ and $I$ (see Bird \& Beers 1993), we 
checked if the distribution is consistent with a single Gaussian
or it is bimodal applying the KMM test (Ashman et al. 1994), using the
program kindly provided by the authors. This test gives the likelihood ratio
between the hypothesis that the dataset is better described by the sum
of two Gaussians and the null hypothesis that the dataset is 
better described by a single Gaussian. 

In Tab. \ref{tab:dynam}, we report
the dynamical parameters for the velocity excesses found in our sample (see
the discussion below). Column (1) refers to the peak identification, 
column (2) reports the number of velocities used, columns (3) and (4)
are the estimated mean velocity and velocity dispersion.

\begin{table}
\caption[]{\label{tab:dynam} Estimated dynamical parameters}
\scriptsize
\begin{flushleft}
\begin{tabular}{llll} 
\hline\noalign{\smallskip}
Peak &  N & $<v>$(km/s)	& $\sigma$ (km/s) \\ \ \cr
\noalign{\smallskip}
\hline\noalign{\smallskip}
Peak A & 64 & 12866$\pm$150 &  869$\pm$ 78	\\		
Clump 1 & 28 & 11951$\pm$116&  267$\pm$59	\\
Clump 2 & 36 & 13498$\pm$111 & 260$\pm$52	\\ 
Peak B & 40 & 30477$\pm$107 & 602$\pm$148	\\
Peak C & 17 & 41603$\pm$ 215 & 849$\pm$217	\\ 
\noalign{\smallskip}
\hline
\end{tabular}
\end{flushleft}
\end{table}

\subsection{Peak A}

\begin{figure}[ht]
\psfig{figure=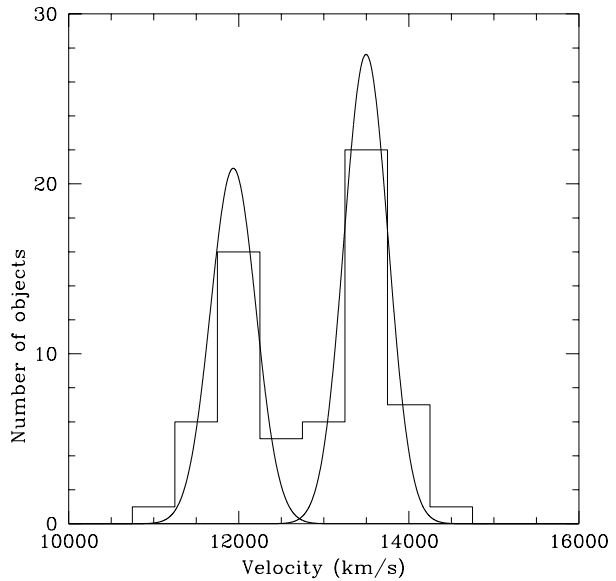,width=8.truecm}
\parbox[b]{7.5cm}{\caption {\label{fig:bipic} {Velocity distribution
of galaxies belonging to peak A with superimposed the two Gaussians
with  $<v>_1=11951$ km/s and $\sigma_1=267$ km/s and $<v>_2=13498$ km/s 
and $\sigma_2=260$ km/s. Among the 64 objects of the peak, 28 have been
assigned to the lower velocity group and 36 to the higher velocity one. 
             }}} 
\end{figure}

Peak A, formed by 64 galaxies, is at the same velocity of A548 and has 
$<v>=12866\pm 150$ km/s and $\sigma=869\pm 78$ km/s. The shape estimators $a$
and  $b_2$ revealed a significant deviation from the null hypothesis
of Gaussianity (at more than $95 \%$ significance level). The KMM test 
revealed that the distribution is significantly better described
by two Gaussians, both in the homoscedastic and in the heteroshedastic case. 
Assigning the objects to the two groups on the basis of the {\it a posteriori}
probability given by the KMM algorithm (see Ashman et al. 1994) and estimating
the dynamical parameters with the biweight estimators, we found 
$<v>_1=11951 \pm 116$ km/s and $\sigma_1=267\pm 59$ km/s (based on 28 
velocities) and  $<v>_2=13498 \pm 111$ km/s and $\sigma_2=260\pm 52$ km/s 
(based on 36 objects). Fig. \ref{fig:bipic} shows a close up of the velocity
distribution of galaxies in peak A, with superimposed the two Gaussians of
parameters $<v>_1$, $\sigma_1$ and $<v>_2$, $\sigma_2$.
No significant differences are found, inside each clump of peak A, 
between the dynamical parameters of galaxies with and without emission lines. 

\begin{figure}[ht]
\psfig{figure=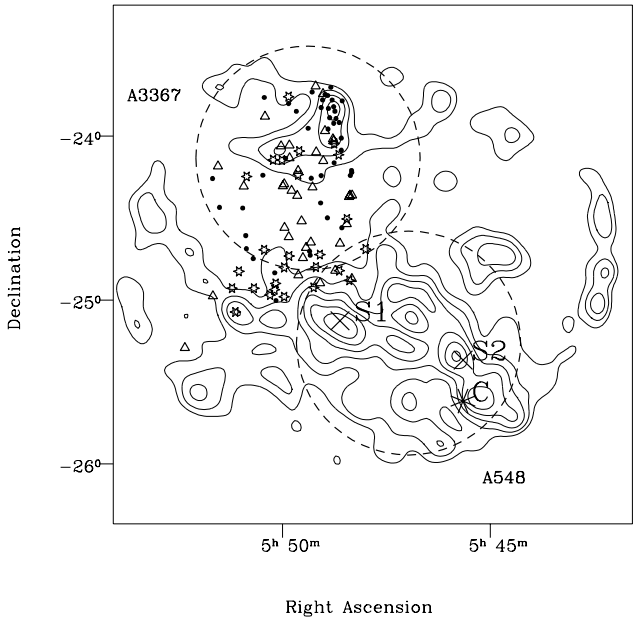,width=9.truecm}
\parbox[b]{7.5cm}{\caption {\label{fig:smoo3} {Positions of galaxies with
$10000 <v<12700$ km/s (stars), with $12700 <v<15000$ km/s 
(triangles) and  $25000 <v<35000$ km/s (black dots).
Crosses refer to the positions of X--ray extended sources and the asterisk
refers to the position of an optical substructure in A548 (see Davis et al. 
1995).
}}} 	 
\end{figure}

These mean velocities can be compared with those
of the subclumps found by Davis et al. (1995) in A548. Our value of $<v>_1$
is well consistent with their Clump $a$ (see their table 4b). Our 
$<v>_2$ is 2.6 $\sigma$ different from the mean velocity of their Clump $b$: 
however assuming that the error on the Davis et al. determination (not 
reported in their paper) could be similar to ours, the discrepancy may be 
reduced to less than 1 $\sigma$. 
Moreover, our values are well in agreement with those reported by Escalera et
al. (1994).

More noticeable is the difference in the velocity dispersions. 
Clumps $a$ and $b$ have values of $638$ km/s and $553$ km/s respectively,
while we estimated $\sim 260$ km/s for both groups. This fact could be an 
indication of a dependence of the velocity dispersion on the distance from
the group centers. 

The dynamical situation of peak A is therefore very similar to that
of A548 and this peak has to be considered an extension of the nearby
cluster rather than a separated entity. In particular, the subcondensation
of peak A at lower velocity is part of Clump $a$, while the higher velocity 
group of peak A is associable to Clump $b$. 

In order to see the relative distance of the clumps detected in A548
from our galaxies, we plotted on the isodensity contours the positions
of their centers and the distribution on the sky of our sample.
In Fig. \ref{fig:smoo3} stars represent galaxies with 
$10000 <v<12700$ km/s and triangles refer to objects with $12700 <v<15000$ 
km/s.  The big crosses are the reported centers of the extended emissions
found in the ROSAT observations labelled as S1 and S2 in table 2 of Davis et al.
and coincident with their optical Clumps $a$ and $b$. 
Their source S3 falls outside the figure. 
We reported also the position of their Clump $c$ as an asterisk.
Note the good coincidence between these positions and peaks in the
density field. 

The extension in A3367 of Clump $a$ of Davis et al. seems to have two 
condensations (see stars in Fig. \ref{fig:smoo3}) 
at $\alpha(2000)\sim 05^h 50^m$, $\delta(2000)\sim -24^o 10'$ 
and $\alpha(2000)\sim 05^h 50^m$, $\delta(2000)\sim -25^o 00'$,
at the distances of $\sim 65$ and $\sim 30$ arcmin from the corresponding 
substructure in A548, respectively. 

The extension in A3367 of Clump $b$ (see triangles in Fig. \ref{fig:smoo3}) 
is more smooth and spans all the distances between $1^o$ and $1.5^o$ from its 
center, corresponding to $\sim$ 2-3 h$^{-1}$ Mpc. 

Finally we note that the mean velocity reported by Postman et al. (1992)
for A3367 is consistent with the redshift of peak A. The velocity of the 
brightest cluster member (Postman \& Lauer 1995) reveals that this galaxy 
is part of the higher velocity clump of peak A: therefore this galaxy is
probably associated to A548 rather than to A3367.

\subsection{Peaks B and C} 

The estimated dynamical parameters of the second peak seen in the
redshift histogram are $<v>=30477\pm 107$ km/s and $\sigma=602 \pm 148$ km/s,
based on 40 velocities. The percentage of emission line galaxies is
$25\%$. In Fig. \ref{fig:smoo3}, the objects belonging
to this structure are plotted as black dots and seem to be concentrated
around a density peak at $\alpha(2000)\sim 05^h 48^m$ and
$\delta(2000)\sim -23^o 50'$.
The shape estimators indicate that the distribution is consistent with being a 
Gaussian. Given the fact that this density excess is the largest one inside
one Abell radius from the nominal center of A3367 and considering that its
velocity dispersion is typical of a cluster, we suggest that the name
A3367 is in fact to be attributed to peak B.

Finally, the estimated mean velocity and velocity dispersion for 
peak C are  $<v>=41603\pm 215$ km/s and $\sigma=849 \pm 217$ km/s, 
determined with 17 objects. The Gaussianity of the distribution can not be 
excluded.

\section{Summary}

We have presented the results of a redshift survey in an area between the
Abell clusters A548 and A3367. These two clusters have been suspected
to be a close pair and therefore candidates to undergo a merging process.

We obtained 180 new galaxy velocities with the use of multifiber spectroscopy,
$44\%$ of which presents emission lines. The redshift histogram shows
clearly three peaks at $v\sim 12000$ km/s, $v\sim 30000$ km/s and 
$v\sim 40000$ km/s: for each of them we have estimated the dynamical parameters.

The first structure (peak A) is at the velocity of A548 and
could be considered as an extension of this clusters. In particular,
this peak is formed by two substructures, corresponding to two
subclusters in A548 revealed by Davis et al. (1995) both in the
optical distribution of galaxies and in a ROSAT X-ray map. 
The velocity dispersions of our two clumps in peak A are significantly smaller
than those determined in the subclumps of Davis et al., while the mean
velocities are in agreement. 

Peak B has $<v>=30477$ km/s and a velocity dispersion
of $\sim 600$ km/s; the distribution of its members on the plane of the sky
corresponds to the highest density excess of A3367: for this reason we
suggest that the name A3367 is in fact to be attributed to this
clump. The mean velocity reported for A3367 by Postman et al. (1992) is based
on galaxies belonging to peak A, and therefore to an extension of A548 rather
than to A3367.

Our general conclusion is that A548 and A3367 is not a close pair of merging 
clusters, being the two structures at significantly different redshift. 
However, we found that the complex dynamical structure of A548 has large 
coherence, with a projected extension in the range of 1-3 h$^{-1}$
Mpc. 

\begin{acknowledgements}
We thank M. Mignoli and D. Maccagni for having observed fields 61 and 62,
and H.B\"ohringer for his help with the photometric catalogue. This research
has made use of the NASA/IPAC Extragalactic Database.
\end{acknowledgements}

\end{document}